# Visible Fluorescence Emission from Self-assembled Porphyrin Nanotubes

Jyotsana Gupta and C. Vijayan

*Department of Physics, Indian Institute of Technology Madras, India*

RECEIVED DATE; E-Mail: cvijayan@iitm.ac.in

Porphyrins form a class of environment-friendly materials available abundantly in nature. Their structure permits scope for molecular engineering in terms of the choice of the central metal ion, the possibility of attaching different ligands to appropriate positions in the molecule and the ability to form aggregates with interesting physical properties [1-3]. Recent work has established the importance of porphyrins as building blocks for self assembled nanostructures with possible applications in a variety of fields in biochemistry, electronics and photonics [4]. These include light harvesting [5], optical waveguide design, optical-frequency conversion [6], nonlinear optical devices [7] as well as information processing, transmission, and storage [8].

Porphyrins have been reported to form strip like structures by J-aggregation [9]. A few papers have come out recently on the design of functional organic nanostructures such as well-formed tubes or fibers based on porphyrins [1-3]. Wang *et al.*, have reported on the synythesis of porphyrin nanotubes (PNTs) by ionic self assembly of two oppositely charged porphyrins in aqueous solution [1]. Kojima *et al.*, have prepared nanoscaled tubular structures via self-assembly using a few template molecules [2]. The investigation of the optical properties of PNTs is an area which has not attracted much attention though these are of crucial importance in the applications being envisaged.

Here we report on the observation of reasonably strong fluorescence emission in the visible region from a newly synthesized PNTs. The PNTs is made out of self assembly of two oppositely charged porphyrins, namely meso-tetrakis (4-sulfonatophenyl) porphyrin dihydrochloride and Fe(III) meso-Tetra (N-Methyl-4-Pyridyl) porphyrin pentachloride. The interesting aspect to note is that the resulting PNT is fluorescent whereas both the parent compounds do not exhibit fluorescence. This is important in view of the need for biofriendly fluorescent materials.

The PNTs are prepared by the following procedure, similar to that used by Wang *et al.*. Solutions of negatively charged meso-tetrakis (4-sulfonatophenyl) porphyrin dihydrochloride in water (10.5 μM, 0.02 M HCl) and positively charged Fe(III) meso-Tetra (N-Methyl-4-Pyridyl) porphyrin pentachloride in water (3.5 μm) are prepared. The two solutions are colored dark green and dark brown respectively. Equal volumes of both solutions are mixed together and kept for 70 hours in dark. room. The resulting solution is yellowish green in color and has a pH value of 2.

The high resolution transmission electron microscopic (HRTEM) image for the PNT obtained by the above process is shown in Figure 1. Long structures of width 100 to 130 nm as well as clusters of tubes of smaller diameter (roughly 20 to 40 nm) attached to the long structures can be seen. (see also the supporting information). Wang *et al*, had reported a reversible transition between rod-like structures and tubes in the case of their PNTs on illumination with light. However, we did not observe any such transition due to irradiation in the present case.

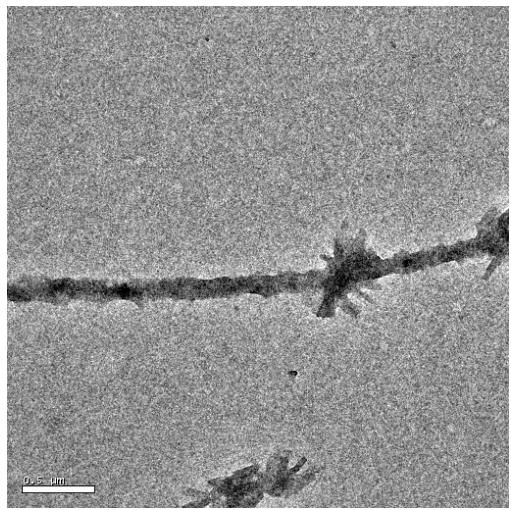

**Figure 1.** HRTEM image of PNT.

The optical absorption spectrum of PNTs shows a peak at 432 nm (Figure 2). The optical density of the PNTs is much less than that of the porphyrin monomers at the spectral regions of their maximum absorption.

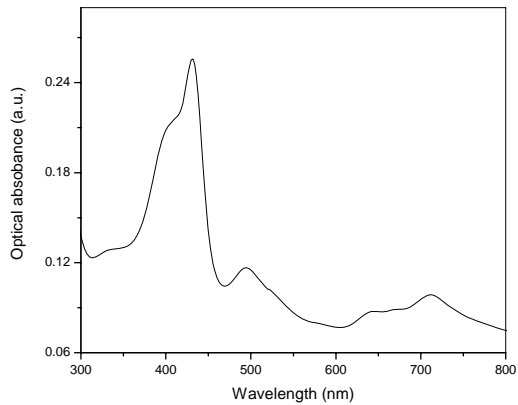

**Figure 2.** Optical absorption spectrum of PNTs.

The fluorescence emission spectrum for excitation at 432 nm is shown in figure 3. The spectrum consists of a well defined single peak at 669 nm. The corresponding excitation spectrum shows a single peak at 432 nm, in the spectral region of the absorption band (Figure S4 of the supporting information). The reddish emission from the sample can be easily seen by unaided eye on excitation at 432 nm.

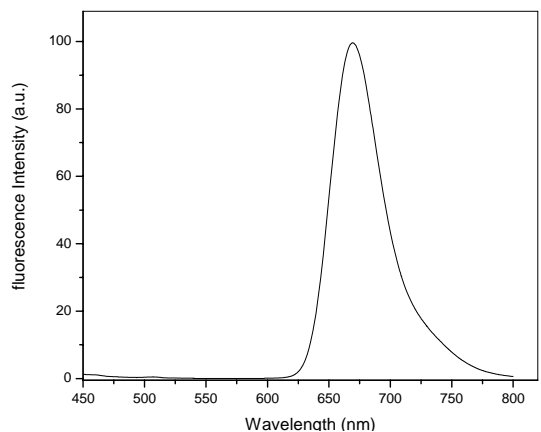

*Figure 3.* Fluorescence emission spectrum of PNTs.

Wang *et al.*, had reported a quenching of the fluorescence in comparison with the porphyrin monomers in the case of their PNTs. However, our samples do exhibit fluorescence eventhough the corresponding porphyrin monomers do not exhibit fluorescence. There have been recent reports on fluorescence emission from different types of nanorod-like structures such as CNTs [10].

Several aspects of formation of nanotubes and other rodlike structures based on porphyrins are yet to be understood properly. Recently Kojima *et al.*, had suggested a mechanism on how curved surfaces could come together and form tubular structures [2]. Liu et al have emphasized the strong influence of the geometry of the metal ion as well as the ligands in the formation of tubular nanostructures [3]. Further studies are being carried out on the mechanism of formation of PNTs and nanorods and on the physical processes responsible for the enhanced fluorescence in comparison with the porphyrin monomers. The observation of fluorescence from biofriendly materials such as pophyrin nanostructures is expected to be of considerable importance in medical applications whereas rendering the material in the form of building blocks for nanostructures offers considerable scope in various other applications of nanotechnology as well.

**Acknowledgment.** We thank Prof. J. Subramanyan from Central Leather Research Institute (CLRI) for fruitful discussion and advice and the Department of Science and Technology Unit on Nanoscience, IIT Madras for help in recording high resolution transmission electron micrographs.

**Supporting Information Available:** Additional experimental details such as HRTEM images optical absorption spectra of porphyrin monomers and fluorescence excitation spectrum of PNTs is available free of charge via the Internet.

# Abstract and artwork

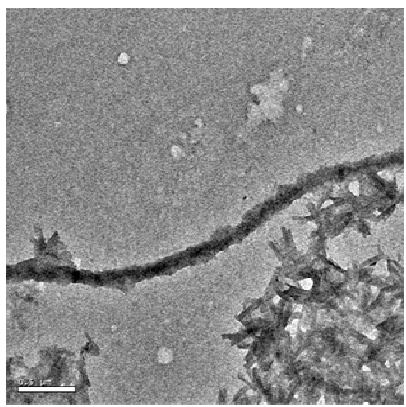 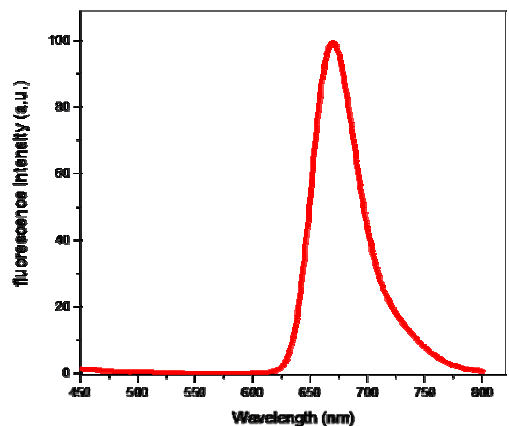

HRTEM image of PNTs     Fluorescence emission from PNTs

Porphyrin nanotubes (PNTs) are prepared by self-assembly using meso-tetrakis (4-sulfonatophenyl) porphyrin and Fe(III) meso-Tetra (N-Methyl-4-Pyridyl) porphyrin in water as starting materials. Long tubes of about one micron diameter are formed with bunches of smaller tubes attached to it, as judged from the analysis of HRTEM images. The PNTs formed by this method are found to exhibit good visible emission at 669 nm on excitation at 432 nm whereas both parent porphyrin monomers do not exhibit any fluorescence. This result highlights the scope of PNTs as functional components in the design of biofriendly devices in medical as well as nanophotonics applications.